\documentstyle[12pt]{article}
\begin{document}
\title{\Large\bf 1996--1998 Polish Visual Meteor Database}
\author{A. Olech$^1$, M. Wi\'sniewski$^2$ and M. Gajos$^2$}
\date{\small $^1$ Nicolaus Copernicus Astronomical Center, ul. Bartycka
18, 00-716 Warszawa, Poland \\
$^2$Warsaw University Observatory, Al. Ujazdowskie 4, 00-478 Warszawa,
Poland}

\maketitle

\begin{abstract} 

The summary of 1996-1998 visual observations collected by the Polish
{\sl Comets and Meteors Workshop} is presented. In total, during 2328.12
effective observing hours 14085 meteors were seen and plotted onto
gnomonic starmaps by 41 observers. The date, time, magnitude, angular
velocity and equatorial coordinates for each observed event are given.
The full 1996--1998 {\sl Polish Visual Meteor Database (PVMDB)} is
accessible via INTERNET 

\end{abstract}

\section{Introduction}

Since 1994 the Polish {\sl Comets and Meteors Workshop (CMW)} has been
cooperating with the {\sl International Meteor Organization (IMO)}.
During the first two years we made mostly visual observations of major
showers without plotting the meteors onto the gnomonic star maps. Over
time the experience of our observers grew and in 1996 we decided to
start visual observations with plotting.

Every year a complete set of our observation reports was sent to the
{\sl IMO} and our results were included in the {\sl IMO Visual Meteor
Database (VMDB)} (see for example Arlt 2000). However we would like to
point out that the {\sl VMDB} contains only the information about hourly
rates and magnitude distributions of the observed meteors. Thus, an
error in classification of a meteor made by the observer while filling
out the report form is included also in the {\sl VMDB}.

Additionally the {\sl VMDB} contains the data only about meteor showers
from the {\sl IMO Working List of the Meteor Showers}. Thus it is
impossible to get the information about other small or poorly known
streams from the {\sl VMDB}.

The solution to the problem is to publish a full database containing all
quantities describing a meteor event including its equatorial
coordinates and angular velocity. Such a database can be searched for
the presence of any shower in any moment of time.

The database of Polish telescopic observations made in the years
1996-1998 was already published by Olech \& Jurek (2000). Following this
approach we decided to publish in the same form our visual results from
the years 1996-1998. Table 1 summarizes our visual work in this period
of time. In a total 14085 meteors were seen during 2328.12 effective
observing hours by 41 observers.

Table 2 shows a list of the {\sl CMW} observers with their effective
observing time and number of meteors plotted in each of years 1996-1998.

\section{Coordinate files}

The files {\tt coor96.txt}, {\tt coor97.txt} and {\tt coor98.txt}
contain data for each observed meteor such as the date of appearance,
serial number of meteor, its magnitude, its angular velocity (in scale
from $A$ to $F$), time of appearance, equatorial coordinates of
beginning and end, IMO code of the observer and three letter code. Below
we show a small sample of such a file: 
\medskip

\leftline{\tt 1998~01~01/02~~~1~~~4.5~C~00:47 219.20~~76.42 237.00~~72.38 SKOAN ABZ}
\leftline{\tt 1998~01~01/02~~~2~~~2.0~B~00:47 321.66~~66.76 005.76~~59.44 SKOAN ABZ}
\leftline{\tt 1998~01~01/02~~~3~~~1.5~C~00:47 216.55~~52.21 236.21~~56.24 SKOAN ABZ}
\leftline{\tt 1998~01~01/02~~~4~~~1.5~C~00:47 257.92~~50.32 266.80~~48.49 SKOAN ABZ}
\leftline{\tt 1998~01~01/02~~~5~~~4.0~D~00:47 211.86~~50.55 206.85~~51.73 SKOAN ABZ}
\leftline{\tt 1998~01~01/02~~~6~~-2.0~B~00:47 097.50~~87.00 312.50~~81.00 SKOAN ABZ}
\leftline{\tt 1998~01~01/02~~~7~~~2.0~B~01:37 206.19~~78.68 251.99~~65.72 SKOAN ACA}
\leftline{\tt 1998~01~01/02~~~8~~~4.0~C~01:37 181.14~~73.42 171.16~~74.95 SKOAN ACA}
\leftline{\tt 1998~01~01/02~~10~~~4.0~D~01:37 273.52~~52.78 269.18~~49.60 SKOAN ACA}
\leftline{\tt 1998~01~02/03~~~1~~~4.5~D~17:01 028.60~~43.07 017.24~~43.14 OLEAR ACB}
\medskip

\noindent and in Table 3 we give byte-by-byte description of these files.

Three letter code shown in the last column of {\tt coor9?.txt} file is
used for connecting each meteor with the information about the
observation stored in the {\tt head9?.txt} file.

The time of appearance of a meteor, when it is not given exactly in the
report form, is assumed as the middle time of each observing period.

All equatorial coordinates were inputed using the {\sc CooReader}
software (Olech \& Samuj{\l}{\l}o 1999).

\section{Header files}

The files {\tt head96.txt}, {\tt head97.txt}, {\tt head98.txt} contain
information about the each observing run such as: three letter code
allowing to connect each observation with data on meteors presented in
coordinate files, IMO code of observer, longitude and latitude of place
of observation, date, UT time of begin and end of observation, solar
longitude (J2000) of middle time of each run, equatorial coordinates of
observed field, effective time of observation, cloud correction factor
$F$, stellar limiting magnitude estimated by the naked eye and the IMO
code of the place of observation.

Below we show a small sample of such a file:
\medskip

\leftline{\tt ABZ SKOAN~~21.0 E 50.0 N 02 01 98 0016 0118 281.444 210~~75 1.00 1.00 5.80 34029}
\leftline{\tt ACA SKOAN~~21.0 E 50.0 N 02 01 98 0118 0156 281.479 210~~75 0.60 1.00 5.72 34029}
\leftline{\tt ACB OLEAR~~23.5 E 51.1 N 02 01 98 1630 1732 282.133 000~~70 1.00 1.00 5.42 34012}
\leftline{\tt ACC OLEAR~~23.5 E 51.1 N 02 01 98 2026 2134 282.302 000~~70 1.00 1.00 5.70 34012}
\leftline{\tt ACD OLEAR~~23.5 E 51.1 N 03 01 98 0005 0108 282.456 000~~70 1.00 1.00 6.18 34012}
\leftline{\tt ACE OLEAR~~23.5 E 51.1 N 03 01 98 0110 0214 282.502 000~~70 1.00 1.00 6.13 34012}
\leftline{\tt ACF OLEAR~~23.5 E 51.1 N 03 01 98 0214 0305 282.543 000~~70 0.75 1.00 6.15 34012}
\leftline{\tt ACG SZAKO~~23.2 E 50.5 N 02 01 98 2003 2124 282.291 181~~53 1.30 1.00 6.40 34040}
\medskip

Table 4 gives a byte-by-byte description of the header files.

\section{Summary}

We have presented the summary of the 1996-1998 visual observations made
by the Polish {\sl Comets and Meteors Workshop}. In total 14085 meteors
were observed during 2328.12 effective observing hours collected by 41
observers. The date, time, magnitude, angular velocity and equatorial
coordinates for each observed event is given. The full 1996--1998 {\sl
Polish Visual Meteor Database (PVMDB)} is accessible via INTERNET at
the following URL: \linebreak {\tt http://www.astrouw.edu.pl/$\sim$olech/VIS/}.

The 1999-2000 data are still under review but they will be available
to the astronomical community as soon as possible.
\smallskip

\noindent {\bf Acknowledgements} We would like to thank to all observers
who sent us their observations. This work was supported by KBN grant 2
P03D 026 20. A.O. acknowledges also support from Fundacja na Rzecz Nauki
Polskiej.

\vspace{2cm}

\begin{table}[h]
\caption{Polish Visual Meteor Database (PVMDB) grand totals for
1996-1998}
\vspace{0.2cm}
\begin{center}
\begin{tabular}{||c|c|r|c||}
\hline
\hline
Year & Observers & $T_{\rm eff}(^h)$ & Meteors \\
\hline
\hline
1996 & 18 &  247.86 & 1508 \\
1997 & 25 &  849.41 & 5269 \\
1998 & 31 & 1230.85 & 7308 \\
\hline
Total & 41 & 2328.12 & 14085 \\
\hline
\hline
\end{tabular}
\end{center}
\end{table}

\begin{center}
\begin{table}[h]
\caption{Total effective observing time in hours ($T_{\rm eff}$) and
number of meteors plotted ($N$) per observer in years 1996-1998.}
\vspace{0.2cm}
{\small
\begin{tabular}{||l|c|r|r|r|r|r|r|r|r||}
\hline
\hline
Observer & IMO & \multicolumn{2}{|c|}{1996} &
\multicolumn{2}{|c|}{1997} & \multicolumn{2}{|c|}{1998} &
\multicolumn{2}{|c|}{Total} \\
\cline{3-10}
         & Code & $T_{\rm eff}$ & $N$ & $T_{\rm eff}$ & $N$
& $T_{\rm eff}$ & $N$ & $T_{\rm eff}$ & $N$\\
\hline
\hline
Jaros{\l}aw Dygos     & DYGJA & ----  & --- &  44.99 & 181 & 308.98 &1324 & 353.97 & 1505\\
Tomasz Fajfer         & FAJTO & 84.50 & 382 & 185.50 & 862 &  22.50 & 115 & 292.50 & 1359\\
Konrad Szaruga        & SZAKO & 26.14 & 144 & 108.15 & 659 &  88.35 & 437 & 222.64 & 1240\\
Krzysztof Socha       & SOCKR & 17.31 & 102 &  87.47 & 616 & 105.11 & 769 & 209.89 & 1487\\
Maciej Kwinta         & KWIMA &  4.67 &  19 &  71.24 & 438 &  68.08 & 540 & 143.99 &  997\\
Gracjan Maciejewski   & MACGR &  ---- &  -- &  49.17 & 219 &  81.17 & 394 & 130.34 &  613\\
Marcin Konopka        & KONMA &  ---- &  -- &  36.39 & 349 &  81.59 & 450 & 117.98 &  799\\
Arkadiusz Olech       & OLEAR & 20.92 & 248 &  42.88 & 540 &  49.75 & 463 & 113.55 & 1251\\
Andrzej Skoczewski    & SKOAN &  ---- &  -- &  46.68 & 276 &  56.84 & 380 & 103.52 &  656\\
Pawe{\l} Trybus       & TRYPA &  ---- &  -- &   2.17 &   8 &  90.55 & 587 &  92.72 &  595\\
Wojciech Jonderko     & JONWO &  2.20 &   5 &  22.17 & 137 &  39.12 & 155 &  63.49 &  297\\
Marcin Gajos          & GAJMR &  6.29 &  37 &  35.17 & 248 &  17.63 & 104 &  59.09 &  389\\
Albert Krzy\'sk\'ow   & KRZAL &  ---- &  -- &  11.83 &  76 &  43.49 & 282 &  55.32 &  358\\
Aleksander Trofimowicz& TROAL &  ---- &  -- &   ---- &  -- &  38.47 & 229 &  38.47 &  229\\
Krzysztof Wtorek      & WTOKR & 23.00 & 140 &  11.99 &  78 &   ---- &  -- &  34.99 &  218\\
{\L}ukasz Raurowicz   & RAULU &  ---- &  -- &  23.62 & 163 &   6.09 &  41 &  29.71 &  204\\
Micha{\l} Jurek       & JURMC &  8.52 &  43 &  14.66 &  93 &   6.00 &  53 &  29.18 &  189\\
Cezary Ga{\l}an       & GALCE &  ---- &  -- &   ---- &  -- &  28.85 & 204 &  28.85 &  204\\
{\L}ukasz Pospieszny  & POSLU & 20.68 & 158 &   6.91 &  30 &   ---- &  -- &  27.59 &  188\\
Luiza Wojciechowska   & WOJLU &  ---- &  -- &   ---- &  -- &  25.32 & 168 &  25.32 &  168\\
Mariusz Wi\'sniewski  & WISMA &  ---- &  -- &   ---- &  -- &  20.86 & 342 &  20.86 &  342\\
Maciej Reszelski      & RESMA &  7.86 &  89 &   8.77 &  99 &   ---- &  -- &  16.63 &  188\\
Pawe{\l} Brewczak     & BREPA &  ---- &  -- &   ---- &  -- &  16.52 &  81 &  16.52 &   81\\ 
{\L}ukasz Sanocki     & SANLU &  5.77 &  39 &   4.34 &  40 &   6.17 &  28 &  16.28 &  107\\
Tomasz Ramza          & RAMTO &  7.00 &  32 &   5.98 &  19 &   ---- &  -- &  12.98 &   51\\
Artur Szaruga         & SZAAR &  ---- &  -- &  10.17 &  37 &   2.12 &   8 &  12.29 &   45\\
Tomasz Dziubi\'nski   & DZITO &  3.50 &  21 &   8.00 &  42 &   ---- &  -- &  11.50 &   63\\
Krzysztof Kami\'nski  & KAMKR &  ---- &  -- &   7.60 &  45 &   1.35 &   8 &   8.95 &   53\\
Jaros{\l}aw Noco\'n   & NOCJA &  ---- &  -- &   ---- &  -- &   6.53 &  21 &   6.53 &   21\\
Waldemar Drozdowski   & DROWA &  ---- &  -- &   1.00 &   3 &   5.40 &  19 &   6.40 &   22\\
Rafa{\l} Kopacki      & KOPRA &  5.50 &  30 &   ---- &  -- &   ---- &  -- &   5.50 &   30\\
Krzysztof Mularczyk   & MULKR &  ---- &  -- &   ---- &  -- &   4.00 &  17 &   4.00 &   17\\
Mariola Czubaszek     & CZUMA &  ---- &  -- &   ---- &  -- &   2.80 &  40 &   2.80 &   40\\
Adam Pisarek          & PISAD &  ---- &  -- &   ---- &  -- &   2.71 &   8 &   2.71 &    8\\
Marek Piotrowski      & PIOMA &  ---- &  -- &   2.56 &  11 &   ---- &  -- &   2.56 &   11\\
Jacek Kluczewski      & KLUJA &  ---- &  -- &   ---- &  -- &   2.00 &  21 &   2.00 &   21\\
Sylwia Che{\l}moniak  & CHESY &  ---- &  -- &   ---- &  -- &   1.50 &  11 &   1.50 &   11\\
Krzysztof Gdula       & GDUKR &  1.50 &   4 &   ---- &  -- &   ---- &  -- &   1.50 &    4\\
Pawe{\l} Musialski    & MUSPA &  1.50 &  11 &   ---- &  -- &   ---- &  -- &   1.50 &   11\\
Sylwia Ho{\l}owacz    & HOLSY &  ---- &  -- &   ---- &  -- &   1.00 &   9 &   1.00 &    9\\
Robert So{\l}tys      & SOLRO &  1.00 &   4 &   ---- &  -- &   ---- &  -- &   1.00 &    4\\
\hline
Total                 & ----- &247.86 &1508 & 849.41 &5269 &1230.85 &7308 & 2328.12&14085\\
\hline
\hline
\end{tabular}}
\end{table}
\end{center}

\clearpage

\begin{table}[t]
\caption{Byte-by-byte description of {\tt coor9?.txt} files}
\vspace{0.2cm}
\begin{center}
\begin{tabular}{|r|c|c|r|}
\hline
\hline
Bytes & Format & Units & Explanations \\
\hline
\hline
1-4    & I4   & - & Year \\
6-7    & I2   & - & Month \\
9-13   & A5   & - & Day/Day \\
15-17  & I3   & - & Number of meteor in report \\
19-21  & F5.1 & mag & magnitude of meteor \\
25     & I1   & - & Velocity in scale form $A$ to $F$ \\
27-31  & A5   & UT & Time \\
33-38  & F6.2 & $^\circ$ & RA of the beginning of meteor (J2000)\\
40-45  & F6.2 & $^\circ$ & Decl. of the beginning of meteor (J2000)\\
47-52  & F6.2 & $^\circ$ & RA of the end of meteor (J2000)\\
54-59  & F6.2 & $^\circ$ & Decl. of the end of meteor (J2000)\\
61-65  & A5   & - & IMO Code of observer \\
67-69  & A3   & - & Three letter code \\
\hline
\hline
\end{tabular}
\end{center}
\end{table}

\begin{table}[d]
\caption[ ]{Byte-by-byte description of {\tt head9?.txt} files}
\vspace{0.2cm}
\begin{center}
\begin{tabular}{|r|c|c|r|}  
\hline
\hline
Bytes & Format & Units & Explanations \\
\hline
\hline
1-3 & A3 & - & Three letter code \\
5-9 & A5 & - & IMO Code of observer \\
11-15 & F5.1 & $\circ$ & Longitude of place of observation \\
17 & A1 & - & Hemisphere designation \\
19-22 & F4.1 & $\circ$ & Latitude of place of observation \\
24 & A1 & - & Hemisphere designation \\
26-27 & I2 & - & Day \\
29-30 & I2 & - & Month \\
32-33 & I2 & - & Year \\
35-38 & I4 & - & Time of beginning of observation (UT) \\
40-43 & I4 & - & Time of end of observation (UT) \\
45-51 & F7.3 & $\circ$ & Solar longitude of middle time of observation (J2000) \\
53-55 & I3 & $\circ$ & RA of center of field of view (J2000) \\
57-59 & I3 & $\circ$ & Decl. of center of field of view (J2000) \\
61-64 & F4.2 & h & Effective time of observation \\
66-69 & F4.2 & - & Cloud correction factor $F$ \\
71-74 & F4.2 & mag & Limiting magnitude estimated in field of view \\
76-80 & I5 & - & IMO code of the place of observation \\
\hline
\hline
\end{tabular}
\end{center}
\end{table}

\end{document}